\documentclass[prb,aps,amsmath,amssymb,showpacs,superscriptaddress, twocolumn]{revtex4-2}
\usepackage[english]{babel}
\usepackage[latin1]{inputenc}
\usepackage{hyperref}
\usepackage{amsmath, amssymb, amsfonts}
\usepackage{amssymb,xcolor}
\usepackage{graphicx}
\usepackage{exscale}
\usepackage{latexsym}
\usepackage{cases}
\usepackage{epstopdf}
\usepackage{subfigure}
\usepackage{bm}
\usepackage{color}

\newcommand{\ket}[1]{|#1\rangle}
\newcommand{\bra}[1]{\langle #1|}
\newcommand{\tr}{\mathrm{Tr}}

\newcommand{\n}{\nonumber\\}

\newcommand{\ex}[1]{\langle #1\rangle}

\newcommand{\overbar}[1]{\mkern 1.5mu\overline{\mkern-1.5mu#1\mkern-1.5mu}\mkern 1.5mu}

\allowdisplaybreaks

\begin{document}

\title{Entanglement and work extraction in the central-spin quantum battery}

\author{Jia-Xuan Liu}
\thanks{These authors contributed equally to this work.}
\address{School of Physics, Northwest University, Xi'an 710127, China}
\address{
Hefei National Laboratory for Physical Sciences at Microscale and Department of Modern Physics,
Universerity of Science and Technology of China, Hefei Anhui 230026,China
}
\author{Hai-Long Shi}
\thanks{These authors contributed equally to this work.}
\address{State Key Laboratory of Magnetic Resonance and Atomic and Molecular Physics, Wuhan Institute of Physics and Mathematics, APM, Chinese Academy of Sciences, Wuhan 430071, China}
\affiliation{University of Chinese Academy of Sciences, Beijing 100049, China}
\author{Yun-Hao Shi}
\address{
	Institute of Physics, Chinese Academy of Sciences, Beijing 100190, China
}
\affiliation{University of Chinese Academy of Sciences, Beijing 100049, China}
\author{Xiao-Hui Wang}
\email{xhwang@nwu.edu.cn}
\address{School of Physics, Northwest University, Xi'an 710127, China}
\address{
Shaanxi Key Laboratory for Theoretical Physics Frontiers, Xi'an 710127, China
	}
\address{
Peng Huanwu Center for Fundamental Theory, Xi'an 710127, China
}
\author{Wen-Li Yang}
\address{
	Institute of Modern Physics, Northwest University, Xi'an 710127, China}
\address{
Shaanxi Key Laboratory for Theoretical Physics Frontiers, Xi'an 710127, China
	}
\address{
Peng Huanwu Center for Fundamental Theory, Xi'an 710127, China
}

\date{\today}

\begin{abstract}
We consider a central-spin battery where $N_b$ central spins serve as battery cells and $N_c$ bath spins serve as charging units.
It is shown that the energy stored in the battery that can be extractable is quantified by the ergotropy, and that  battery-charger entanglement is quantified via the Von Neumann entropy.
By using an exact approach to a one-cell and two-cell battery, our analytical results suggest that, during the charging process, the extractable work slowly increases before the battery-charger entanglement reaches its maximum and then it will rapidly increase  when the entanglement begins to decrease.
In particular, we rigorously show that there is an inverse relationship between the extractable work and the entanglement at the end of the charging process.
Moreover, we investigate different approaches to realize optimal work extraction without wasted energy.
Among them a central-spin battery with an unpolarized Dicke state as the charger possesses a universal charging time $\propto 1/N_c$, large extractable work, and $\sqrt{N_c}$-improvement of charging power compared with the  battery in the Tavis-Cummings limit.
The above-mentioned results have also been numerically verified in multi-cell batteries.
Our results pave the way to improve extractable work storage in the central-spin battery and highlight a competitive relation between the extractable work and the battery-charger entanglement.
\end{abstract}

\maketitle
\section{Introduction}
The state-of-the-art technology of qubit manipulation enables us to exploit quantum resource, such as entanglement or coherence, for technological purposes \cite{Nielsen,Bennett96,Giovannetti04,Gisin02,Bennett93,Lloyd96,DiVincenzo95}.
A recent development of this direction is studying ``quantum batteries'', which harness the unique property of quantum thermodynamics to speed up the charging process and extract more work compared to their classical counterparts.
The ideal of quantum batteries was first put forward by Alicki and Fannes in 2013 \cite{Alicki13}. 
They demonstrated that entangling unitary controls (i.e., collective controls)  perform better than individual controls (i.e., parallel controls) in work extraction.
Further research uncovered that entanglement generation benefits the speedup of work extraction \cite{Hovhannisyan13}.
Subsequently, the authors of Refs. \cite{Binder15,Campaioli17} argued that, in the charging process, the collective charging scheme results in $k$ time faster than the parallel charging scheme, where $k$ denotes a $k$-body interaction among battery cells.
Therefore, we now say quantum batteries have quantum advantage.

The above-mentioned works mainly focused on abstract time-evolution operators.
To realize such operators in practice, especially the collective controls, various theoretical models have been considered \cite{Ferraro18,Rossini20,Rosa20,Kamin20,Andolina18,Andolina19,Andolina19-2,Peng21,Barra19,Le18,Ghosh20,Carrega20,citekey20,Pirmoradian19,Zhang19,Farina19,Xie18,Zhang18,Quach20,Crescente20-2}.
For instances, the spin-chain battery uses an intrinsic spin-spin interaction to realize the collective controls \cite{Le18} while the Dicke battery relies on the cavity photons to generate an effective interaction among its battery cells \cite{Ferraro18,Zhang18}.
Based on superconducting qubits, a quantum battery has  just recently been realized in experiment \cite{Hu21}.

\begin{figure}[t]
\includegraphics[width=2.5in]{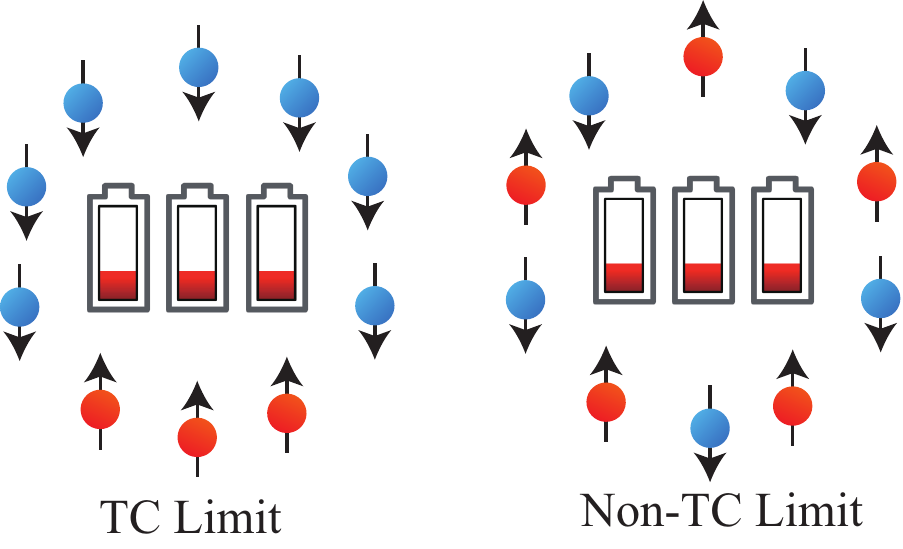}
\caption{Schematic illustration of the central-spin battery.
Central spins serve as $N_b$ battery cells whereas bath spins serve as $N_c$ charging units.
%
%
In the limit of $N_c\to\infty$, if the number of spin-up charging units is far less than $N_c$ then the central-spin battery reduces to the TC battery and we call this case the TC limit, see the left panel.
On the other hand, if the number of spin-up charging units is proportional to $N_c$ but not zero or one then the correspondence between the central-spin battery and the TC battery breaks and we call this case the non-TC limit, see the right panel.
}\label{fig1}
\end{figure}

The central spin model, as an exactly solvable model, has played a vital  role in quantitatively understanding decoherence problem and entanglement dynamics \cite{Schliemann2003,Quan2006,Hanson2008,Bortz10,Barnes12,He19,Lu20,Wan20,Claeys18,Wu20}.
This model can be  experimentally realized by superconductors \cite{Faribault19}, quantum dots \cite{Yao2006,Sarma2009,Faribault2013}, and nitrogen-vacancy centers in diamond \cite{Doherty2013}.
Quantum battery problem can be naturally described in this model and we call it the ``central-spin (quantum) battery'', where $N_b$ central spins serve as  ``battery cells'' and $N_c$ bath spins serve as   ``charging units'', see Fig. \ref{fig1}.
However,  research on the central-spin battery remains limited \cite{Peng21}.
Assuming that only few  photons in the view of the Holstein-Primakoff transformation, the central-spin battery will reduce to the Tavis-Cummings (TC) battery, which is a simplification of the Dicke battery.
The charging performance of the central-spin battery in this limit has been confirmed to have quantum advantage \cite{Peng21}.
A recent Ref. \cite{Andolina19} pointed out that battery-charger entanglement is a major obstacle in work extraction by comparing the performance of a classical charger (coherent state) with that of nonclassical chargers (Fock and squeezed states) in the TC battery.
However, the correspondence between the central-spin battery and the TC battery breaks when the number of photons is far greater than the number of charging units. 

To bridge this gap, we fix our attention on the central-spin battery, especially the non-TC case, and discuss the following questions.
Is there a quantitative relationship between  battery-charger entanglement and  extractable work?
Under what conditions  ``optimal'' work extraction can be achieved? 
The ``optimal'' refers to that all energy stored in the battery can be extracted.

The paper is organized as follows.
In Sec. \ref{S-CSB}, we introduce the central-spin battery and discuss its correspondence with the TC battery.
Time evolutions for  one-cell battery and two-cell battery are analytically solved in the same section.
Multi-cell battery cases are treated by a numerical method. 
With the above preparation, we obtain  analytical and numerical results about the evolutions of  battery-charger entanglement and  battery's extractable work in Sec. \ref{S-Ent-Work}. 
%
%
We rigorously show that the battery-charger entanglement first increases and then decreases during the charging process.
At the end of the charging process, the final  extractable work  is significantly enslaved to the battery-charger entanglement.
To be more exact, the extractable work increases with the decrease of the battery-charger entanglement when the initial chargers are restrained to the Dicke states.
In Sec. \ref{S-Optimal-Work}, we show that  optimal work extraction can be realized when the central-spin battery lies in the non-TC limit region (e.g. the initial charger is an unpolarized Dicke state).  
Moreover, in this region, the charging time is proportional to $1/N_c$, which is independent of the number of battery cells. 
On the other hand, if the central-spin battery is in the TC limit, the optimal work extraction is not always possible unless the number of photons is large but does not break the TC limit.
At this case, the charging time is  proportional to $1/\sqrt{N_c}$.
Finally, a conclusion is given in Sec. \ref{S-Con}.
Our analytical results shed light on  entanglement and extractable work in the central-spin battery and provide different proposals able to optimally store extractable work.

\section{Central-spin battery}\label{S-CSB}
The central-spin battery, just as its name implies, is governed by the Hamiltonian of the central spin model,
\begin{eqnarray}\label{CSB}
&&H=H_b+H_c+H_I,\\ 
&&H_b=BS^z,\quad H_c=hJ^z,\n
&&H_I=A(S^+J^-+S^-J^+)+2\Delta S^z J^z,\nonumber
\end{eqnarray} 
where $S^{\alpha}=\sum_{j=1}^{N_b}\sigma_j^\alpha/2$ $(\alpha=x,y,z)$ are the total spin operators for $N_b$ battery cells and $J^{\alpha}=\sum_{k=1}^{N_c}\sigma_k^\alpha/2$ for $N_c$ charging units.
$J^\pm=J^x\pm iJ^y$ and $S^\pm=S^x\pm iS^y$ are spin ladder operators.
The Hamiltonians of the battery and the charger are denoted by $H_b$ and $H_c$, respectively.
The parameter $A$ and $\Delta$  characterize the flip-flop interaction and the Ising interaction, respectively.

 At time $t<0$, the battery is prepared in the ground state of $H_b$, i.e., all spins are down $\ket{0}_b\equiv\ket{\downarrow_1,\downarrow_2,\ldots,\downarrow_{N_b}}$.   
 The charger is prepared in a Dicke state with $m$ spins up
 \begin{eqnarray}\label{Dicke state}
 \ket{m}_c\equiv\frac{1}{\sqrt{\binom{N_c}{m}}}\sum_{k}P_k(\ket{\uparrow_1,\cdots,\uparrow_m,\downarrow_{m+1},\ldots,\downarrow_{N_c}})
 \end{eqnarray}
 where $\binom{N_c}{m}=N_c!/[m!(N_c-m)!]$ and $P_k$ denotes the complete set of all possible distinct permutations of the qubits.

At time $t=0$, by suddenly turning on the interaction Hamiltonian $H_I$ for a finite time $T$, we aim to inject as much as possible energy into the battery.
Such a  time $T$ is called the charging time.
%
%
To evaluate the performance of a quantum battery, we need obtain the reduced density matrix for battery cells, which is given by $\rho_b(t)=\tr_c(e^{-iHt}\rho_0e^{iHt})$, where $\rho_0=\ket{\psi_0}\bra{\psi_0}$ is the initial state and $\ket{\psi_0}=\ket{0}_b\otimes\ket{m}_c$.
To ensure that $\ket{0}_b$ is the ground state of $H_b=BS^z$, we assume $B>0$ in the following discussion. 

\subsection{Two limits }

Before calculating $\rho_b(t)$ we first discuss the relationship between our battery and the TC battery.
The Holstein-Primakoff (HP) transformation establishes an exact map from the total spin operators $J^\alpha$ to a bosonic mode $a$,
\begin{eqnarray}\label{HP-T}
&&J^+\to\sqrt{N_c}a^\dag\sqrt{1-\frac{a^\dag a}{N_c}},\\
&&J^-\to\sqrt{N_c}\sqrt{1-\frac{a^\dag a}{N_c}}a,\\
&&J^z\to-\frac{N_c}{2}+a^\dag a.
\end{eqnarray}
Note that the Dicke state $\ket{m}_c$ is an eigenstates of $J^z$ with the eigenvalue $(-N_c/2+m)$. 
Thus, the number of up bath spins is exactly the number of photons in the view of the HP transformation.
For convenience we will identify  up spins with photons.
If the number of photons is far less than $N_b$ then the HP transformation simplifies to $J^+\to \sqrt{N_c} a^\dag, J^-\to\sqrt{N_c}a$ and the central-spin battery (\ref{CSB}) reduces to the TC battery, i.e.,
\begin{eqnarray}
H&=&BS^z+h\left(-\frac{N_c}{2}+a^\dag a\right)\\
& &+A\sqrt{N_c}(S^+a+S^-a^\dag)+2\Delta S^z\left(-\frac{N_c}{2}+a^\dag a\right).\nonumber
\end{eqnarray} 
We call this limit 
\begin{eqnarray}\label{TCLimit}
\lim_{N_c\to\infty}\frac{\ex{a^\dag a}}{N_c}=
\lim_{N_c\to\infty}\frac{m}{N_c}=0
\end{eqnarray}
the TC limit, where $m$ is the number of up bath spins or the number of photons.

Another case is that the number of photons is proportional to $N_c$ but not zero or one, i.e.,
\begin{eqnarray}\label{non-TC}
\lim_{N_c\to\infty}\frac{\ex{a^\dag a}}{N_c}=\lim_{N_c\to\infty}\frac{m}{N_c}\equiv k\neq 0,1.
\end{eqnarray} 
We call it the non-TC limit.
In this limit, the correspondence between the central-spin battery and the TC battery breaks due to the non-negligible term $\sqrt{1-a^\dag a/N_c}$ in the HP transformation (\ref{HP-T}).
Theses two limits (\ref{TCLimit}) and (\ref{non-TC}) are illustrated in Fig. (\ref{fig1}).

\subsection{Exact solutions}
Now we evaluate the reduced density matrix of the battery.
Without loss of generality, we assume that the number of up-spin charging units is not less than the number of battery cells, i.e., $m\geq N_b$.
Due to the $u(1)$ symmetry, namely, $[H,J^z+S^z]=0$, the invariance subspace of $H$ containing the initial state $\ket{\psi_0}=\ket{0}_b\otimes\ket{m}_c$ is given by 
\begin{eqnarray}\label{Invariant-subspace}
\mathcal H_m=\span\{\ket{0}_b\ket{m}_c,
\ket{1}_b\ket{m-1}_c,\cdots,
\ket{N_b}_b\ket{m-N_b}_c\},\n
\end{eqnarray}  
where both of the states of battery and charger are expressed in term of the Dicke state, see (\ref{Dicke state}).
For example, $\ket{2}_b\ket{3}_c$ refers to the quantum state whose battery part is a Dicke state with 2 particles and whose charger part is a Dicke state with 3 particles.

In terms of the basis (\ref{Invariant-subspace}), the Hamiltonian (\ref{CSB}) can be represented as a $(N_b+1)\times(N_b+1)$ matrix:
\begin{eqnarray}\label{N-1}
\bm H=\begin{pmatrix}
b_0&u_1&\\
u_1&b_1&u_2&\\
&\ddots&\ddots&\ddots&\\
&&u_{N_b-1}&b_{N_b-1}&u_{N_b}\\
&&&u_{N_b}&b_{N_b}
\end{pmatrix},
\end{eqnarray} 
 where $u_j=A\sqrt{j(N_b-j+1)(N_c-m+j)(m-j+1)}$ and $b_j=B(j-N_b/2)+h(m-j-N_c/2)+2\Delta(j-N_b/2)(m-j-N_c/2)$.
Suppose that $\bm H$ is diagonalized by a unitary matrix $\bm U$, that is $\bm H=\bm U \bm D\bm U^\dag$ where $\bm D$ is a diagonal matrix.
Then, the matrix representation of the wavefunction for the whole system at time $t$ is given by
\begin{eqnarray}\label{N-2}
\bm{\psi}(t)=\bm{U} e^{-i\bm Dt}\bm U^\dag (1\ 0\ \ldots\ 0)^T,
\end{eqnarray}   
and 
\begin{eqnarray}\label{total-state}
\ket{\psi(t)}=\bm\psi_1(t)\ket{0}_b\ket{m}_c+\cdots+\bm\psi_{N_b+1}(t)\ket{N_b}_b\ket{m-N_b}_c.\n
\end{eqnarray}
It thus follows that the reduced density matrix of the battery is given by 
\begin{eqnarray}\label{N-3}
\rho_b(t)&=&\tr_c(\ket{\psi(t)}\bra{\psi(t)})\n &=&|\bm \psi_1(t)|^2\ket{0}\bra{0}+\cdots+|\bm\psi_{N_b+1}|^2\ket{N_b}\bra{N_b}.
\end{eqnarray}
Eqs. (\ref{N-1}, \ref{N-2}, \ref{N-3}) give a numerical approach to evaluate $\rho_b(t)$ in multi-cell battery cases.

For the one-cell battery case $N_b=1$, we find that $\bm U=\exp(-i\theta\sigma^y/2)$ and $\bm D={\rm diag}(d_1,d_2)$, where $\theta$ is determined by
\begin{eqnarray}\label{theta}
&&\sin\theta=\frac{u_1}{\sqrt{u_1^2+\frac{1}{4}(b_0-b_1)^2}},\n
&&\cos\theta=\frac{b_0-b_1}{2\sqrt{u_1^2+\frac{1}{4}(b_0-b_1)^2}},
\end{eqnarray}
and the energy levels are given by
\begin{eqnarray}
&&d_1=\frac{b_0+b_1}{2}+\sqrt{u_1^2+\frac{1}{4}(b_0-b_1)^2},\\
&&d_2=\frac{b_0+b_1}{2}-\sqrt{u_1^2+\frac{1}{4}(b_0-b_1)^2}.
\end{eqnarray}
The reduced density matrix of the battery is thus given by
\begin{eqnarray}\label{rho-1}
\rho_b(t)=\frac{1+r(t)}{2}\ket{0}\bra{0}+\frac{1-r(t)}{2}\ket{1}\bra{1},
\end{eqnarray}
where 
\begin{eqnarray}\label{rho-1-r}
r(t)=\cos^2\theta+\cos((d_1-d_2)t)\sin^2\theta.
\end{eqnarray}

For the two-cell battery case $N_b=2$, we let $h=B$ and $\Delta=0$ to simplify the calculation.
Under this assumption, the matrices $\bm U$ and $\bm D$ are given by
\begin{eqnarray}
&&\bm U=\frac{1}{\sqrt{2(u_1^2+u_2^2)}}\begin{pmatrix}
\sqrt{2}u_2& u_1&u_1\\
0&\sqrt{u_1^2+u_2^2}&-\sqrt{u_1^2+u_2^2}\\
-\sqrt{2}u_1&u_2&u_2
\end{pmatrix},\n
\end{eqnarray}
and $\bm D={\rm diag}(e_1,e_2,e_3)$, where $u_1$ and $u_2$ are the same as the ones in Eq. (\ref{N-1}) and 
\begin{eqnarray}
&&e_1=B(m-1-N_c/2),\\
&&e_2=B(m-1-N_c/2)+\sqrt{u_1^2+u_2^2},\\
&&e_3=B(m-1-N_c/2)-\sqrt{u_1^2+u_2^2}.
\end{eqnarray}
The reduced density matrix of the battery is given by
\begin{eqnarray}
\rho_b(t)=\rho_{11}(t)\ket{0}\bra{0}+\rho_{22}(t)\ket{1}\bra{1}+\rho_{33}(t)\ket{2}\bra{2},
\end{eqnarray}
where 
\begin{eqnarray}\label{rho-elements}
\rho_{11}(t)&=&\frac{1}{(u_1^2+u_2^2)^2}(
u_2^2+u_1^2\cos(\omega t))^2,\n
\rho_{22}(t)&=&\frac{u_1^2}{u_1^2+u_2^2}(1-\cos^2(\omega t)),\n 
\rho_{33}(t)&=&\frac{u_1^2u_2^2}{(u_1^2+u_2^2)^2}(1-\cos(\omega t))^2,
\end{eqnarray}
and $\omega=e_2-e_1=\sqrt{u_1^2+u_2^2}$.

\section{Entanglement and extractable work}\label{S-Ent-Work}
\begin{figure*}[t]
\includegraphics[width=6in]{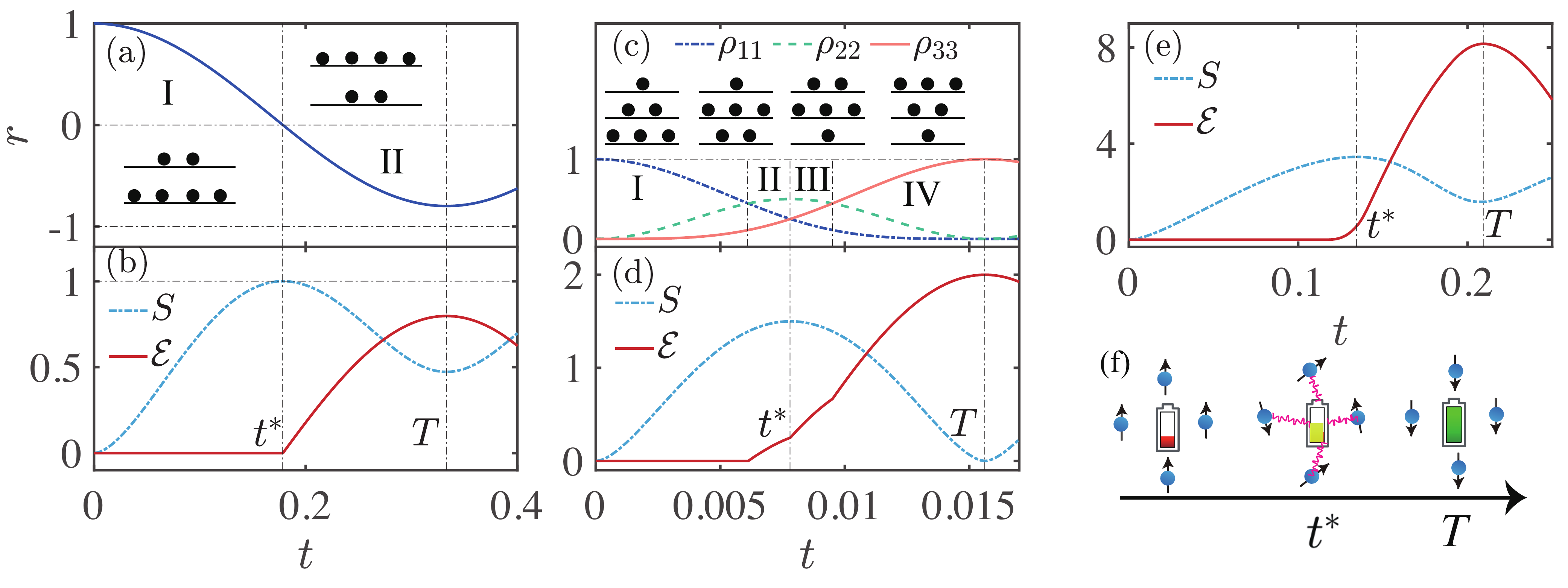}
\caption{
Evolution of the occupation numbers for the battery part are plotted in panel (a) for $N_b=1$ and panel (c) for $N_b=2$.
Evolutions of battery-charger entanglement $S(t)$ and extractable work $\mathcal E(t)$ are plotted in panel (b) for $N_b=1$, panel (d) for $N_b=2$ and panel (e) for $N_b=10$.
$t^*$  denotes the time when the battery-charger entanglement reaches its maximum.
$T$ denotes the time when the extractable work reaches its maximum.
The process of $0\sim T$ is called the charging process.
The cartoon (f) illustrates the changes of battery-charger entanglement during the charging process.
The other parameters  are set to be 
$B=1, h=4, A=1,\Delta=0,N_b=1,N_c=20, m=20$ in panels (a,b); 
$B=h=1, , A=1,\Delta=0,N_b=2,N_c=200, m=100$ in panels (c,d); 
$B=h=1, , A=1,\Delta=0,N_b=10,N_c=20, m=20$ in panel (e).
}\label{fig2}
\end{figure*}
Since the quantum state of battery-charger system remains a pure state during the time evolution, the battery-charger entanglement can be well characterized by the Von Neumann
entropy of the battery part, which is given by \cite{Nielsen}
\begin{eqnarray}
S(t)=-\tr(\rho_b(t)\log_2\rho_b(t)).
\end{eqnarray}
Another important quantity is the extractable work.
During the charging process, the energy injected from the charger to the battery is described by
\begin{eqnarray}\label{energy}
\Delta E(t)=E(t)-E(0),
\end{eqnarray}
where $E(t)=\tr(\rho(t)H_b)$ denotes the mean energy of the battery state $\rho_b(t)$.
However, not all energy can be extracted from the battery, which is known as the second law of thermodynamics.
A proper measure of the extractable work  for the state $\rho_b(t)$ is provided by the ergotropy \cite{Allahverdyan04}  
\begin{eqnarray}\label{ergotropy}
\mathcal E(t)=E(t)-E_p(t),
\end{eqnarray}
where $E_p(t)=\tr(\overbar{\rho_b(t)}H_b)$ is the energy of the passive counterpart $\overbar{\rho_b(t)}$ of $\rho_b(t)$.
Passive states are incapable of giving out energy via any cyclic Hamiltonian process.
In our settings, the passive state $\overbar{\rho_b(t)}$ is given by
\begin{eqnarray}
\overbar{\rho_b(t)}=\sum_{j=1}^{N_b+1}|\bm\psi_{\eta_j}|^2\ket{j-1}\bra{j-1},
\end{eqnarray}  
where $\bm\psi$ is given in Eq. (\ref{N-3}) and $\eta$ is a permutation of $1,2,\ldots, N_b+1$ so that $|\bm \psi_{\eta_{j}}|^2\geq |\bm \psi_{\eta_{j+1}}|^2$ for $j=1,\ldots, N_b$.

Now we consider the single-cell battery case, i.e., $N_b=1$.
The entanglement for this case reduces to the binary entropy \begin{eqnarray}\label{Nb-1-S}
S(t)\!=\!-\frac{1+r(t)}{2}\log_2\frac{1+r(t)}{2}-\frac{1-r(t)}{2}\log_2\frac{1-r(t)}{2},\n
\end{eqnarray}
which shows an inverse relationship between $S(t)$ and $|r(t)|$.
However, the ergotropy is
\begin{eqnarray}\label{erg-Nb-1}
\mathcal E(t)=\left\{
\begin{array}{lrr}
0,&&r(t)\geq 0\\
-Br(t),&&r(t)< 0\\
\end{array}
\right.
\end{eqnarray}
which is positively related to the $|r(t)|$ when $\mathcal E(t)\neq 0$.
Thus, we have the following 
\\ \emph{Theorem1}:
For $N_b=1$, if the ergotropy $\mathcal E(t)$ is not equal zero then $\mathcal E(t)$ is in inversely related to the battery-charger entanglement $S(t)$.
In particular, at the end of the charging process, $\mathcal E(T)$ is also in inversely related to $S(T)$.

As shown in Figs. \ref{fig2}(a) and \ref{fig2}(b), $r(t)$ keeps falling in the time interval $0\sim T$.
At time $t^*$, the occupation numbers of the ground state $\ket{0}$ and the excited state $\ket{1}$ are balanced and thus the battery state is maximally entangled with the charger.
When time is far away from $t^*$, the occupation numbers begin to reverse resulting in the decrease of entanglement and the appearance of nonzero extractable work.
At time $T$, the charging process is finished with the most energy stored in the battery and the battery-charger entanglement decreases to a minimum.
According to theorem 1, the extracted work $\mathcal E(T)$ essentially depend on the battery-charger entanglement $S(T)$.
The smaller $S(T)$ the larger $\mathcal E(T)$.

For $N_b=2$ case, the complex expression of Eq. (\ref{rho-elements}) impedes the further discussion.
We notice that $u_1=A\sqrt{2m(N_c-m+1)}$ can be considered as the same as $u_2=A\sqrt{2(N_c-m+2)(m-1)}$ under some special limits, e.g., the Non-TC limit (\ref{non-TC}).
Then Eq. (\ref{rho-elements}) immediately reduces to a simple form:
\begin{eqnarray}\label{simplify}
\rho_{11}&=&
\frac{1}{4}(1+\cos(\omega t))^2,\n
\rho_{22}&=&\frac{1}{2}(1-\cos^2(\omega t)),\n 
\rho_{33}&=&\frac{1}{4}(1-\cos(\omega t))^2.
\end{eqnarray} 
The corresponding extractable work and entanglement are thus given by
\begin{eqnarray}
\frac{\mathcal E(t)}{B}&=&\left\{
\begin{array}{lrr}
0,&&0\leq t< t_1\\
-\frac{3}{4}\left(\cos(\omega t)+\frac{1}{3}\right)^2+\frac{1}{3},&&t_1\leq t< t^*\\
-\frac{3}{4}\left(\cos(\omega t)+1\right)^2+1,&&t^*\leq t< t_2\\
-2\cos(\omega t),&&t_2\leq t\leq T\\
\end{array}
\right.
\end{eqnarray} 
and 
\begin{eqnarray}\label{S-Nb-2-u1=u2}
S(t)&=&\frac{3}{2} +\frac{1}{2}\cos^2(\omega t)-\frac{1}{2}(\cos(\omega t)+1)^2\log_2(\cos(\omega t)+1)\n 
& &-\frac{1}{2}(\cos(\omega t)-1)^2\log_2(1-\cos(\omega t))\n 
& &-\frac{1}{2}(1-\cos^2(\omega t))\log_2(1-\cos^2(\omega t)),
\end{eqnarray}
where $t_1=\arccos(1/3)/\omega, t_2=\arccos(-1/3)/\omega, t^*=\pi/(2\omega),$ and $T=\pi/\omega$.
The analytical results are consistent with the numerical results obtained in the settings of $N_c=200, m=100$, see Figs. \ref{fig2}(c) and \ref{fig2}(d).
Unlike the $N_b=1$ case, the extractable work $\mathcal E(t)$ is not always zero but still small before the entanglement $E(t)$ reaches its maximum $E(t^*)$
because at time $t_1$ the occupation numbers of the ground state and the first excited state are starting to inverse.
Another difference lies in that $S(T)\neq 0$ for $N_b=1$ while $S(T)=0$ for $N_b=2$.
Note that the choice of parameters $N_b,N_c,m$ largely determines whether $S(T)$ is zero.
In the next section, we will discuss this question in detail  and further highlight a strong link between the entanglement and the extractable work.
Now we claim that  
\\ \emph{Theorem2}:
For $N_b=2$, the ergotropy $\mathcal E(T)$  is in inversely related to the battery-charger entanglement $S(T)$.
Here, we need not make the  simplification  by requiring the condition $u_1=u_2$.
\\ \emph{Proof:} 
The charging time $T$ is determined by maximizing the energy injected to the battery, i.e.,  $\Delta E(t)$.
According to the general expressions of the occupation numbers (\ref{rho-elements}) and the definition (\ref{energy}), we have 
\begin{eqnarray}
\Delta E(t)&=&B+\frac{B}{(u_1^2+u_2^2)^2}[
u_1^2(u_2^2-u_1^2)x^2-4u_1^2u_2^2x\n 
& &+u_2^2(u_1^2-u_2^2)],
\end{eqnarray}
where $x=\cos(\omega t)$.
If $u_2^2-u_1^2\geq 0$ then $\Delta E(t)$ takes the maximum value when $x=-1$.
Otherwise, the maximum value of the quadratic function $\Delta E(t)$ is taken at $x_*=-2u_2^2/(u_1^2-u_2^2)$ by ignoring the constraint $-1\leq x\leq 1$.
It is not difficult to show that $x_*<-1$ by using the condition $m\geq N_b=2$.
Thus,
we deduce that the charging time $T$ is determined by $\cos(\omega T)=-1$.
It immediately follows that the occupation number of the first excited state $\rho_{22}(T)=0$.
Therefore, this theorem can be proven by using the same argument in proving theorem 1.

\begin{figure}[t]
\includegraphics[width=3in]{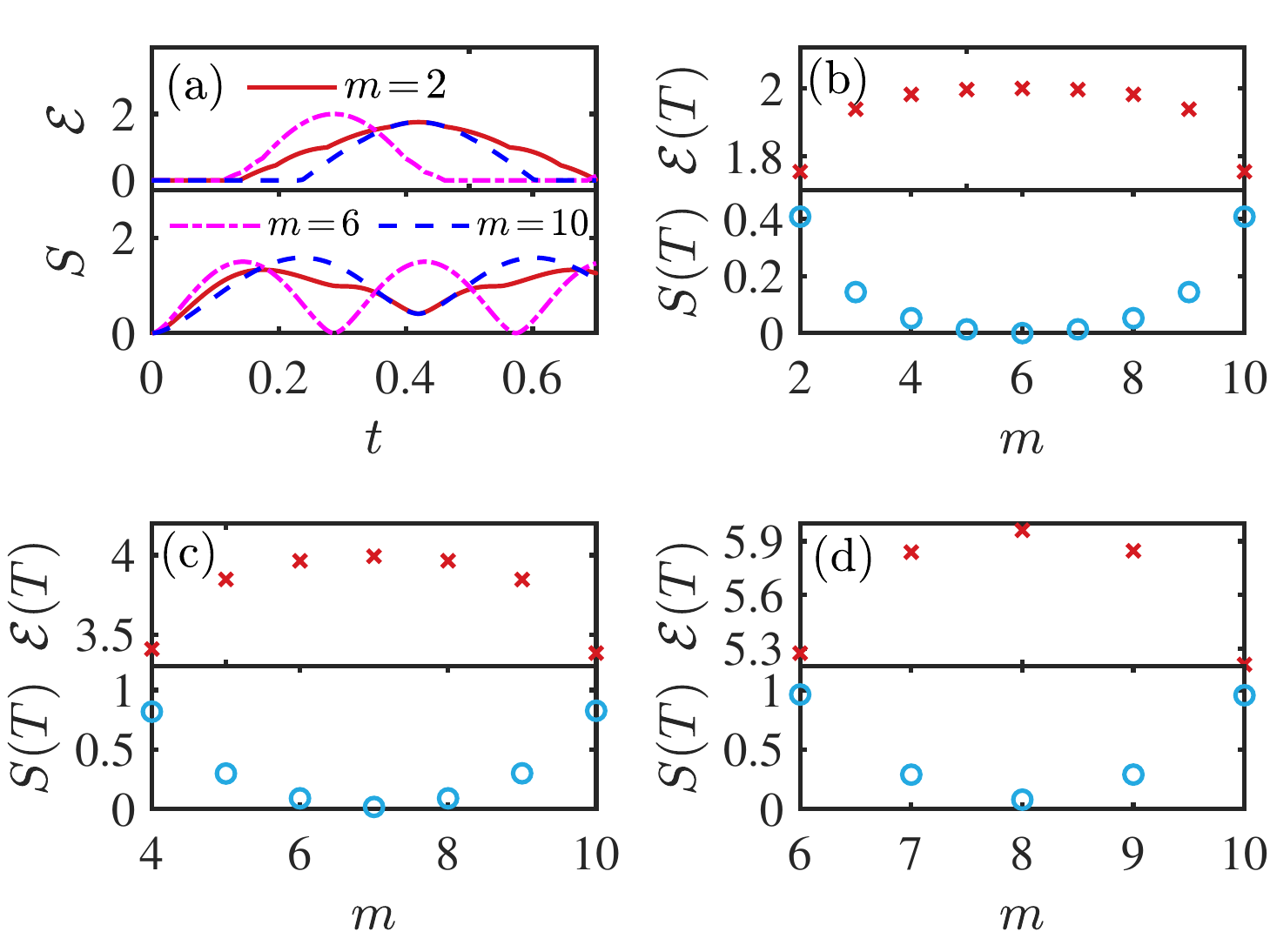}
\caption{
Panel (a) plots the battery-charger entanglement $S(t)$ and the extractable work $\mathcal E(t)$ for $N_b=2, N_c=10$.
From panel (a) we can determine $T$ at which  $\mathcal E(t)$ reaches its maximum $\mathcal E(T)$.
Panels (b-d) plots $\mathcal E(T)$ and $S(T)$ versus $m$.
$m$ is the number of spin-up  charging units. 
Here we set $h=B=A=1, \Delta=0,N_c=10$ for panels (a-d), $N_b=2$ for panels (a,b), $N_b=4$ for panel (c), and $N_b=6$ for panel (d).
}\label{fig3}
\end{figure}

Except for the cases of $N_b=1$ and $N_b=2$, it is difficult to carry out analytical calculations in multi-cell battery cases.
Thus, we perform a numerical verification of the results obtained above for the case of $N_b=10$, see Fig. \ref{fig2}(e).
We see that the minimal value of entanglement $E(T)$ marks the maximal  extractable work $\mathcal E(T)$.
The inverse relationship between $S(T)$ and $\mathcal E(T)$ is checked in Fig. \ref{fig3}.

The cartoon \ref{fig2}(f) shows a variation of the battery-charger entanglement during the charging process.
Initially, the spin-down battery cells are interacted with a well prepared charger that is a Dicke state with $m$ up spins.  
Thus, there is no entanglement between the battery and the charger.
When the charging process begins, the spin down battery cells will be excited due to the flip-flop interaction, resulting in increasing  battery-charger entanglement.
At time $t^*$, the entanglement reaches its maximum and the occupation numbers for the battery display a nearly balanced distribution.
As the charging process continues, the battery cells will continue to be excited while the battery-charger entanglement begins to decrease.
At the end of the charging process, namely time $T$, the energy of battery reaches its maximum  while the battery-charger entanglement reaches its minimum.
However, the extractable work $\mathcal E(T)$ is less than the energy stored in the battery
and is inversely related to the battery-charger entanglement $S(T)$.

\section{Optimal work Extraction}\label{S-Optimal-Work}
In Sec. \ref{S-Ent-Work} we establish an inverse relationship between the battery-charger entanglement $S(T)$ and the extractable work $\mathcal E(T)$, where $T$ is the charging time.
In this section, we will discuss how to realize the optimal work extraction.
Review the definitions given in Eqs. (\ref{energy}) and (\ref{ergotropy}).
Generally, the extractable work $\mathcal E(T)$ is less than the energy $\Delta E(T)$ stored in the battery  since we can not extract any work from the passive state.
Thus, ``optimal'' in here refers to that $\mathcal E(T)=\Delta E(T)$, or equivalently, $E_p(T)=E(0)$.
For convenience, we can let the initial energy $E(0)=0$ by subtracting a constant $E(0)$ from the Hamiltonian (\ref{CSB}).
Then, the condition of optimal work extraction can be stated as that at the end of the charging process, the energy of passive state is zero, i.e., $E_p(T)=0$.
Another important observation is that vanishing battery-charger entanglement $S(T)=0$ implies that  the battery $\rho_b(T)$ is a pure state.
Then, the corresponding passive state is exactly the initial state $\ket{0}_b$ and thus $E_p(T)=E(0)=0$.
Therefore, the battery-charger entanglement $S(T)$ also indicates the quality of extractable work.
The zero entanglement means that the battery possesses optimal extractable work.

For the $N_b=1$ case, we deduce from Eqs. (\ref{rho-1-r}, \ref{erg-Nb-1}) that $T$ is determined by $\cos((d_1-d_2)T)=-1$.
Thus, we have 
\begin{eqnarray}\label{Nb-1-erg_T}
\mathcal E (T)=B\left[1-\frac{2(h-B)^2}{4A^2m(N_c-m+1)+(B-h)^2}\right],
\end{eqnarray}
where $m$ is the number of spin-up charging units and we assume here  $\Delta=0$.
Eq. (\ref{Nb-1-erg_T}) indicates that extractable work $\mathcal E(T)$ decreases when $m$ is far away $(N_c+1)/2$.
According to Eqs. (\ref{theta}, \ref{rho-1-r}, \ref{Nb-1-S}) the entanglement $S(T)=0$ for $m=(N_c+1)/2$ when $h=B$ and $\Delta=0$.

For the $N_b=2$ case, the charging time $T$ is determined by $\cos(\omega T)=-1$.
Substituting this condition into Eq. (\ref{rho-elements}) and the definition of ergotropy (\ref{ergotropy}), we obtain
\begin{eqnarray}
\mathcal E(T)=2B\left[-1+\frac{8}{\frac{u_1^2}{u_2^2}+\frac{u_2^2}{u_1^2}+2}\right].
\end{eqnarray}
It is clear that as $u_1/u_2=\sqrt{m(N_c-m+1)/[(m-1)(N_c-m+2)]}$ deviates from 1, or equivalently, $m$ deviates from $(N_c+2)/2$, the extractable work $\mathcal E(T)$ will decrease, see Fig. \ref{fig3}(b).
The expression (\ref{S-Nb-2-u1=u2}) of entanglement for the case of $m=(N_c+2)/2$ shows that $S(T)=0$ and thus optimal work extraction can be realized.

\begin{figure}[t]
\includegraphics[width=3in]{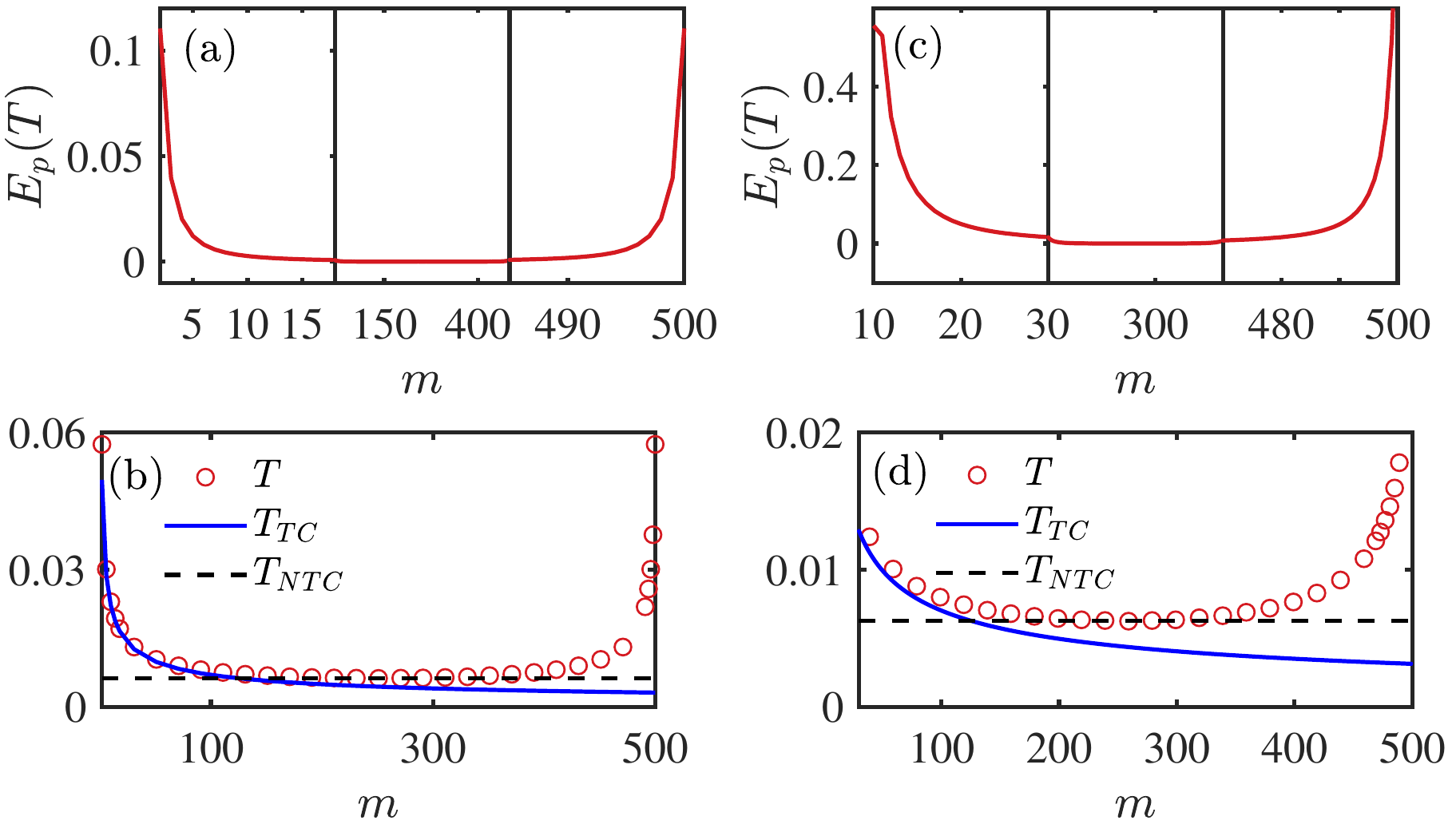}
\caption{
The energy $E_p(T)$ (a,c) of passive state at time $T$ versus the number $m$ of charging units with initially spin up.
$T$ is the charging time.
$T_{NTC}$ and $T_{TC}$ are obtained from Eqs. (\ref{T-NTC-f},\ref{T-TC}).
Here, we set $N_b=2, N_c=500, h=B=A=1,\Delta=0$ for panels (a,b) and $N_b=10,N_c=500,h=10,B=A=1,\Delta=0$ for panels (c,d). 
}\label{fig4}
\end{figure}

The above analytical analysis implies a conjecture that for a fixed $N_b$, the extractable work $\mathcal E(T)$ decreases as $m$ deviates from $(N_c+N_b)/2$. 
Numerical calculations (Figs. \ref{fig3}(b-d)) support this conjecture
.
However, such maximal extractable work may not be optimal.
As shown in Fig. \ref{fig3}(d), $S(T)$ is not zero for $m=(N_b+N_c)/2=8$.

Whether optimal work extraction can be achieved under $N_c\to\infty$?
Taking $N_b=2$ as an instructive example, the condition $u_1=u_2$ ensures that the battery-charger entanglement $S(T)$ (\ref{S-Nb-2-u1=u2}) becomes zero at the end of the charging process.
In Sec. \ref{S-Ent-Work} we have discussed that the condition $u_1=u_2$ can be realized under the non-TC limit.
In the TC-limit (\ref{TCLimit}) an additional condition $m\to \infty$ is required to ensure $u_1=u_2$.
Although these two limits ensure the realization of optimal work extraction, the corresponding charging times are indeed different.
In the non-TC limit, the charging time is given by
\begin{eqnarray}\label{T-NTC}
T_{NTC}=\frac{\pi}{2A\sqrt{k(1-k)}N_c},
\end{eqnarray}  
where $k=\lim_{N_c\to\infty}m/N_c$ is given in the non-TC limit (\ref{non-TC}).
Since $m=(N_b+N_c)/2$ benefits the work extraction, we take $k=1/2$ in  Eq. (\ref{non-TC}) and thus
\begin{eqnarray}\label{T-NTC-f}
T_{NTC}=\frac{\pi}{AN_c}.
\end{eqnarray} 
In the TC limit, the charging time is given by
\begin{eqnarray}\label{T-TC}
T_{TC}=\frac{\pi}{2A\sqrt{mN_c}},
\end{eqnarray}
which indicates the charging power is $\propto \sqrt{N_c}$.
Such $\sqrt{N_c}$-speedup of charging has also been founded in Ref. \cite{Peng21}.
Compared with the performance of the battery in the TC limit, however, the $\sqrt{N_c}$-improvement of charing power in the non-TC limit manifests the uniqueness of the central-spin battery distinguished from the TC battery.

Figs. \ref{fig4}(a,b) plot the energy of passive state and the charging time $T$ for $N_b=2,N_c=500$.
With the increase of $m$, the central-spin battery firstly behaves like the TC battery since the charging time $T$ corresponds to $T_{TC}$, see Fig. \ref{fig4}(b).
Only when $m$ is large enough ($>10$ in Fig. \ref{fig4}(a)) can the energy of passive state becomes zero, so optimal work extraction can be achieved.
With the further increase of $m$, the central-spin battery enters to the non-TC limit and the charging time $T$ conforms to $T_{NTC}$, see Fig. \ref{fig4}(b).
It is worth noting that the central-spin battery behaves like under the non-TC limit for a wide range of parameter $m$ (from 100 to 400 in Fig. \ref{fig4}(b)).
The case of $N_b=10, B\neq h$ is plotted in Figs. \ref{fig4}(c,d).
In Fig. \ref{fig4}(d), the inconsistency between $T_{TC}$ and $T$ arises since $T_{TC}$ (Eq. (\ref{T-TC})) is derived for $N_b=2$ case.
However, $T_{NTC}$ is still consistent with the $T$ when $m$ is around $(N_c+N_b)/2$ even if $h\neq B$.
In this sense, $T_{NTC}$ is universal and independent of the number of quantum cells $N_b$ and the strength of magnetic fields $B$ and $h$.

\section{Conclusion}\label{S-Con}
In this work, we have studied extractable work and entanglement in the central-spin battery.
It is shown that the extractable work is quantified by the ergotropy and the battery-charger entanglement is quantified by the Von Neumann
entropy.
Exact dynamics for them have been obtained in the cases of single-battery cell and two-battery cells.
Using these exact expressions, we rigorously show that during the charging process the battery-charger entanglement first increases to the maximum which marks a nearly  balanced distribution of occupation numbers,
and then entanglement decreases with the rapid increase of extractable work.
At the end of charging process, the final extractable work from battery is in inversely related to the battery-charger entanglement.
The optimal extractable work, namely, no wasted energy, can be achieved if there is no battery-charger entanglement at the end of charging process.
We show that there are two conditions to obtain optimal extractable work: one is the TC limit with $m\to\infty$ and another is the non-TC limit.
For the former, the central-spin battery reduces to the TC battery with the charging time $T_{TC}\propto1/\sqrt{N_c}$, while the latter, the non-TC limit, is unique for the central-spin battery and does not correspond to the TC battery.
We can realize the non-TC limit by preparing the charger in an unpolarized Dicke state.
The advantage of this setting is short charging time $T_{NTC}=\pi/(AN_c)$, large extractable work, and realization of optimal work extraction.
Moreover,  $T_{NTC}$ is independent of the number of battery cells and thus universal.
The above-mentioned analytical results have also been verified by numerical calculations for a ten-cell battery.
It is worth noting that our present results are applicable to quantum batteries where no coherence is involved during the charging process.
Our work deepens the understanding of entanglement and extractable work in such incoherent quantum batteries
and sheds light on how to realize optimal work extraction in the central-spin battery. 
For a more general situation, other quantum correlations will be involved and affect the performance of work extraction.
The relations between extractable work and other quantum correlations will becomes subtle and complicated. 
Uncovering their relations in more general settings will become important.

\section*{ACKNOWLEDGMENTS}
Qing-Kun Wan and Shu Ding are acknowledged for useful discussion.
This work was supported by NSFC (Grants Nos. 12047502, 11975183 and 11875220), the Key Innovative Research Team of Quantum Many-body theory and Quantum Control in Shaanxi Province (Grant No. 2017KCT-12), the Major Basic Research Program of Natural Science of Shaanxi Province (Grant No. 2017ZDJC-32), and the Double First-Class University Construction Project of Northwest University.

H.-L. Shi and J.-X. Liu  contributed equally to the
numerical and analytical studies for this research.

\end{document}